%% file: paper.tex
\documentclass[11pt]{article}

\usepackage{amsmath}
\usepackage{amsfonts}
\usepackage{amssymb}
\usepackage{latexsym}
\usepackage{graphics}
\usepackage{epsfig}
\usepackage{color}
\usepackage{cite}

\input{texdefs}

\begin{document}

\thispagestyle{empty}

\begin{flushright}
\end{flushright}
\vskip 2 cm
\begin{center}
{\Large {\bf 
One-loop Self Energies in Softly Broken Supersymmetric QED 
}
}
\\[0pt]
\vspace{1.23cm}
{\large
{\bf Tino S. Nyawelo\footnote{
{{ {\ {\ {\ E-mail: tnyawelo@physics.utah.edu}}}}}}}, 
{\bf Paolo Gondolo\footnote{
{{ {\ {\ {\ E-mail: paolo@physics.utah.edu}}}}}}},
\bigskip }\\[0pt]
\vspace{0.23cm} 
{\it  
Department of Physics, University of Utah,\\ 
115 S 1400 E, \# 201, 
Salt Lake City, UT 84112-0830, USA.
}
\bigskip
\vspace{.5cm} 
\end{center}
\subsection*{\centering Abstract}

We apply the supergraph with spurion technique to compute the renormalization of one-loop diagrams that contributes to electron and selectron self energies in a softly broken Supersymmetric Quantum Electrodynamics (SQED). In particular, we calculate the one-loop gauge superfield contribution to the two-point function.

\newpage 


\setcounter{page}{1}

\section{Introduction} 
\label{sc:intro} 
Supersymmetric theories have been always been famous for their extraordinary renormailzation properties. One of them being the absence of
quadratic divergences. It is this property that has led to the
development of the Minimal Supersymmetric Standard Model (MSSM).

It is known for a long time that if supersymmetry is realized in nature it must be broken because no super multiplet has been observed. Models of spontaneously broken supersymmetry were constructed  \cite{Fayet:1974jb} and their renormalization was studied extensively \cite{Piguet:1980fy,Sibold:1980mr}. 

Gauge theories with softly broken supersymmetry have been widely studied. 
Almost all of the parameters of the Minimal Supersymmetric Standard Model (MSSM) encode possible  ways in which supersymmetry is broken in a soft way, while preserving the gauge symmetries of the MSSM. To break supersymmetry without destroying 
the renormalization properties of globally supersymmetric  theories, in particular the non-renormalization theorems and the cancellation of quadratic divergencies, 
one has to introduce soft terms. The soft parameters have essentially been 
classified in by Girardello and Grisaru~\cite{Girardello:1981wz}. These authors gave a list of possible soft breaking term  with the new logarithmic divergences they generate: They are either gaugino masses, Hermitian and complex mass matrices for the
complex scalars or three--linear scalar couplings. These masses
can either be introduced on the component level of the supersymmetric
theory, or described by spurion insertions, $\gth^2$ and
$\bgth{}^2$. The latter approach is very powerful because it leaves
most of the supersymmetric structure intact. In this paper, we describe supersymmetry breaking terms in a way that takes most
advantage of the special properties of supersymmetric theories. 

In previous publication \cite{Nibbelink:2006si} we investigated the effective action of (softly) broken supersymmetric theories at the one loop level. We focused on the renormalization of soft parameters in a softly broken supersymmetric 
models. Since our one loop expressions for \Kh\ potential and for the soft parameters are complicated, it is not easy to extract the renormalization information such as the wave function renormalization $Z\,$ and  masses renormalization $Z_m$.  In this paper we perform supergraph computations of the one-loop diagrams that contributes to electron and selectron self energies in globally softly broken Supersymmetric Quantum Electrodynamics (SQED) from which one can read off these renrmalization constants. We employ again the standard supergraph techniques that can be found in the textbooks \cite{Gates:1983nr,West:1990tg}, and use spurions $\gth^2$ and $\bgth^2$ to parameterize soft supersymmetry breaking terms.  We find that our one-loop result generates divergent terms involving spurions $\gth^2$ and $\bgth{}^2$ \cite{Girardello:1981wz,Chang:1985qd,Yamada:1994id}, which we removed by Yamada spurion dependent field redefinition \cite{Yamada:1994id}. We follow the convention\footnote{Our conventions for Gamma matrices are slightly different from those of \cite{Wess:1992cp}: \(\gg^m ~=~ \pmtrx{0 &i\gs^m \\i\bgs^m &0}\), and \(\gg_5 ~=~ \pmtrx{\Id_2 &0 \\ 0 &-\Id_2}\).
} of textbook by Wess and Bagger \cite{Wess:1992cp}.

The plan of the paper is as follows: In sections \ref{sc:sqed} and \ref{sc:spropa} we introduce the classical action of the extended model of the SQED including the soft supersymmetry breaking and calculate the associated superpropagators. The computation of the one-loop diagrams that contributes to electron and selectron is performed in section \ref{s:example}. In subsection \ref{sc:seselectronMass}, we first consider the one loop self energy due to the soft selectron mass insertions, becuase it easier than the ones that result from soft photino mass insertions. Its contribution is described in details in \ref{sc:sephotinoMass}. In both cases we find that the divergent part of the wave function renormalization is independent of the soft breaking, and is the same for selectron and the electron. In section \ref{sc:counterterms}, we introduce counter terms that are needed to cancel divergences we encountered in these examples.

Finally, there are two appendices. Appendix \ref{sc:idendities}, gives some relations which we used in \ref{sc:sephotinoMass} to simplify spurion operators when they are sandwiched between two superspace delta functions $\gd_{21}  \,=\, \gd^4(x_2-x_1) \, \gd^4(\gth_2-\gth_1)\,$, as naturally happens in the evaluation one loop supergraphs. Appendix~\ref{sc:basicintegrals}, describes the regularization by dimensional reduction of the divergent integrals we encountered in section \ref{s:example}.
\section{Super Quantum Electrodynamics} 
\label{sc:sqed} 
In this section we consider globally Super Quantum Electrodynamics with  soft  supersymmetry breaking interactions. The theory of Super Quantum Electrodynamics consists of two oppositely
charged chiral multiplets $\gF_+ = (\varphi_+, \gps_+, F_+)$ 
and $\gF_- = (\varphi_-, \gps_-, F_-)$ 
under a $\U{1}$ gauge
symmetry of which $V = (A_m, \gl^\ga, \bgl^\dga, \cD)$ 
is the vector superfield.

The components of these superfields for the vector multiplet $V$ in the Wess--Zumino 
gauge and the chiral multiplets $\gF_\pm$  are identified by (The vertical bar $|$ at the end of an expression indicates that we have set all $\gth_\ga = \bgth_\dga = 0\,.$)
\begin{eqnarray}
\varphi_\pm &~=~& \gF_\pm |~,~~\qquad\bar{\varphi}_\pm ~=~ \bgF_\pm |~,~~\qquad F_\pm ~=~ \frac{D^2}{-4}\gF_\pm |~,~~\qquad\bF_\pm ~=~ \frac{\bD^2}{-4}\bgF_\pm |~,\non\\
[2ex]
\gps^\ga_\pm &~=~& \frac{D^\ga}{\sqrt{2}}\gF_\pm |~,\quad\bgps^\dga_\pm ~=~ \frac{\bD^\dga}{\sqrt{2}}\bgF_\pm |~,\quad\gl_\ga ~=~ - \frac{i}{4}D_\ga\bD^2 V|~,\quad\bgl_\dga ~=~  \frac{i}{4}\bD_\dga D^2 V|~, \non\\
[2ex]
\gs^m_{\ga\dga} A_m &~=~& - \frac{1}{2}[D_\ga, \bD_\dga] V|~,\qquad ~~~~~\bD^2D^2 V| ~=~ 8 (\cD + i \der_m A^m)~.
\end{eqnarray}
Here ($\cD, F_\pm$) are auxiliary scalar fields,  $\varphi_+$ and $\varphi_-$ are respectively left-handed and right-handed selectron fields\footnote{The notations $\varphi_+\equiv\varphi_L$,  and $\varphi_-\equiv\varphi^*_R$ are current in the literature, but not always convenient in displaying equations.}; and  $A_m$ is a real vector field, the photon. The complex Weyl spinors ($\bgps^\dga_\pm,\gps^\ga_\pm$) and ($\bgl^\dga,\gl^\ga$) are combined to form one massive Dirac spinor, the electron and a photino Majorana spinor respectively (see  \eqref{Dirac}). 

The supersymmetric action for the SQED theory is given by 
\begin{eqnarray}
S_{\rm susy}~=~ \dsp \int \d^8 z\,\Bigl( 
\bgF_+ \,e^{+2eV}\,\gF_+ + \bgF_- \,e^{-2eV}\,\gF_-\Bigl)
\,+\, 
\Bigl\{\int \d^6 z\, \Big(   
m\, \gF_+ \gF_- 
\,+\, \frac 1{4} 
\, W^\ga\,W_\ga
\Big)
\,+\, \text{h.c.}\Bigl\}~,
\label{Acsusy}
\end{eqnarray}
where we use the full and chiral superspace measures, 
$\d^8 z \,=\, \d^4x \, \d^4\gth$ and 
$\d^6 z \,=\, \d^4x \, \d^2\gth\,$, respectively, $m$ is the mass of the electron superfield and $e$ is the $\U{1}$ charge. In this action we
have introduced the Abelian superfield strength   
\equ{
W_\ga ~=~ 
- \frac 14\, \bD{}^2 D_\ga V~, 
}
with the components
\begin{eqnarray}
W_\ga | &~=~& - i \gl_\ga~,\quad D_\gb W_\ga |~=~ - i (\gs^{mn}\ge)_{\gb\ga}F_{mn} - \ge_{\gb\ga}\cD~,\quad D^2W_\ga |~=~ - 4 \gs^m_{\ga\dga}\der_m\bgl^\dga~,\non\\
[2ex]
\bW_\dga | &~=~&  i \bgl_\dga~,\quad~~\bD_\gb \bW_\dga |~=~ - i (\ge\bgs^{mn})_{\dgb\dga}F_{mn} - \ge_{\dgb\dga}\cD~,\quad \bD^2\bW_\dga |~=~ - 4 \der_m\gl^\ga\gs^m_{\ga\dga}~.
\end{eqnarray}
A Fayet--Iliopoulos term
can be included, but we have not done so here. 

To include soft  supersymmetry breaking interactions we extend
this theory by including the following soft action 
\equ{
S_{\rm soft} ~= ~ 
\int \d^6 z\, \gth^2 \Big(   
M^2 \, \gF_+ \gF_- 
\,+\, \frac 1{2} m_\tgg\, \,W^\ga\,W_\ga
\Big) 
~+~ \text{h.c.}~, 
\label{Acsoft}
}
where $M$ is a complex scalar
mass, and $m_\tgg$ is the photino mass. The factor in front of the
photino mass \(m_\tgg\) has been chosen such that the normalization of the
kinetic term of the gaugino is taken into account.

The component form of the full SQED action,  after eliminating the auxiliary fields $(\cD, F_\pm)$ and their complex congugates reads 
\begin{eqnarray}
S_{\rm full} &~=~& S_{\rm susy} + S_{\rm soft}\non\\
[2ex]
 &~=~& \int \d^4 x\,\Bigl[- \frac{1}{4}F_{mn}F^{mn} - \frac{1}{2}\bar{\tgg}\Big(\der\slashed - \tm_\tgg\Big)\tgg- \Bigl|\der_m\varphi_\pm\,\mp\,i e A_m\varphi_\pm\Bigl|^2 \,-\, \bgPs_D\Bigl(\der\slashed -  i e A\slashed + m\Bigl)\gPs_D\non\\
[2ex]
&&
 \,-\, m^2\,(\bar{\varphi}_+\varphi_+ + \bar{\varphi}_-\varphi_-) + M^2 \, (\varphi_+ \varphi_- + \bar{\varphi}_+ \bar{\varphi}_-)  - \frac{1}{2} e^2(\bar{\varphi}_+\varphi_+ - \bar{\varphi}_-\varphi_-)^2\non\\
[2ex]
&&
- \,i\, \sqrt{2}\,e\,\Bigl(\varphi_+\bgPs_DP_R\tgg + \bar{\varphi}_+\bar{\tgg}P_L\gPs_D - \bar{\varphi}_-\bar{\tgg}P_R\gPs_D - \varphi_-\bgPs_DP_R\tgg\Bigl)\Bigl]~,
\label{Acsusycom}
\end{eqnarray}
where the notation $\pm$ indicates that we sum over $+$ and $-\,$. In this expression, we have introduced the electron Dirac spinor and the photino Majorana spinor
\equ{
\gPs_D~=~\pmtrx{\gps_{+\ga} \\ \bgps_-^\dga} ~=~ \gPs_{+L} + \gPs_{-R}~,
\quad \tgg~=~\pmtrx{\gl_\ga \\ \bgl^\dga}~,\quad P_{L}~=~ \frac12(\Id_4 + \gg_5)~,\quad P_{R}~=~\frac12(\Id_4 - \gg_5)~.
\label{Dirac}
}

\section{The Superpropagators} 
\label{sc:spropa} 
After the strictly classical discussion we now turn towards the quantization of the theory using path integral methods. To this end we need to determine the propagators of the superfields $\gF_+$, $\gF_-$, $V$ by coupling them to the sources $J_+$, $J_-$, $J_V$ respectively. Because of the super gauge invariance the kinetic
operator of the vector multiplet is not invertible. This requires 
gauge fixing and the introduction of the corresponding supersymmetric Fadeev--Popov  ghosts $C,C', \bC, \bC'\,$ (see e.g.\
\cite{Gates:1983nr,West:1990tg}). Since the gauge superfields appear
in the same way as in the supersymmetry preserving theory, we use the gauge fixing action \cite{Ovrut:1981wa} (see
also~\cite{Nibbelink:2005wc}) 
\equ{
S_{\rm G.F.} ~=~ - 
\,\int \d^8z\, \gTh\, \bgTh~, 
\qquad 
\gTh ~=~ \sqrt 2\, \frac {\bD{}^2}{-4} V
\label{AcGF}
}
as if supersymmetry is unbroken. This implies that
the FP--ghost sector is the same as in the supersymmetric theory in
the Feynman--'t~Hooft gauge~\cite{Ovrut:1981wa,Nibbelink:2005wc}. 

By using the vector notation 
\equ{
\gF ~=~ \pmtrx{\gF_+ \\ \gF_-}~,
\qquad 
\bgF ~=~ \pmtrx{\bgF_+ & \bgF_-}~, 
\qquad 
J ~=~ \pmtrx{J_+ \\ J_-}~,
\qquad 
\bJ ~=~ \pmtrx{\bJ_+ & \bJ_-}~,
}
for the electron superfields and the chiral sources,  we can write the
quadratic chiral and vector superfields action after gauge fixing as
\equa{ \dsp 
S_{\rm quad} ~=
&
\, \dsp  \int\d^8z\, \Big\{ 
\bgF\,P_+\,\gF 
\dsp 
\,+\, \Big(\frac 12\, \gF^T \Big[ 
m\, P_- + M^2\, \get_- 
\Big] \frac{D^2}{-4 \, \Box}\, \gF 
\,+\, J^T \frac{D^2}{-4 \, \Box}\, \gF ~+~ \text{h.c.}\Big)
\non \\[2ex] &  
~~~~~~~~~~~~~~~\dsp - 
\,V\Big(\, \Box
\,-\, m_\tgg\,  \,\Box^{1/2}\,\get_V 
\,-\, m_\tgg\, \,\Box^{1/2}\, \bget_V \,-\, J_V\,
\Big)V\Big\}~. 
\label{AcgF2}
}
Here we have made use of the chiral projection operators
\equ{
P_+ ~=~ \frac{\bD{}^2 D^2}{16 \,\Box}~, 
\qquad 
P_- ~=~ \frac{D^2 \bD{}^2}{16 \,\Box}~, 
}
and spurion operators 
\begin{eqnarray}
\get_\pm ~=~ \, P_\pm \, \gth^2 \, P_\pm~,
\qquad 
\bget_\pm ~=~ \, P_\pm \, \bgth^2 \, P_\pm~,\qquad 
\get_V ~=~ \frac{D^\ga \, \gth^2\, \bD{}^2 D_\ga }{8  \,\Box^{1/2}}~, 
\qquad 
\bget_V ~=~ \frac{\bD_\dga \, \bgth^2\, D^2 \bD{}^\dga }{8  \,\Box^{1/2}}~.
\labl{chprojec}
\end{eqnarray}
Furthermore, we have used the fact that 
\(
P_+\,\gF = \gF,~P_-\,\bgF = \bgF\)
to rewrite the integral as a full superspace integral. From the quadratic superfield actions \eqref{AcgF2} we determine the propagators for chiral and vector superfields. 

We begin with the chiral multiplet. The propagators of chiral multiplet are obtained from the action \eqref{AcgF2} by rewriting the quadratic action for the chiral superfields as
\equ{
S_\gF ~=~ \int\d^8z\,\Big\{\frac 12\,
\pmtrx{ \gPs^T & \bgPs } 
\,\gD^{-1}\,
\pmtrx{\gPs \\ \bgPs^T}\,+\, \pmtrx{ \bJ & J^T } 
\,\pmtrx{\gPs \\ \bgPs^T}\Big\}~, 
}
using the field redefinitions 
\equ{
\gF^T = \frac{\bD^2}{-4}\,\bgPs~,\qquad \bgF^T = \frac{D^2}{-4}\,\gPs~.
}
Here the superscripts $T$ denotes transposition, $\gPs(\bgPs)$ is (anti--)chiral, and the quadratic operator $\gD^{-1}$ is given by
\equ{
\gD^{-1} ~=~ \pmtrx{(m\, + M^2\, \bgth^2)\frac{D^2}{-4}  & \Box\,P_-\\[1ex] \Box\,P_+& (m\, + M^2\, \gth^2)\frac{\bD^2}{-4} }~.
} 
The path integral can be evaluated in the usual way and we find that the functional integral 
\equ{
Z_0(J,\bJ) ~=~ \int  \cD \gF \cD \bgF \, e^{i \, S_\gF} ~=~  \text{exp}\,\Big\{- \frac i 2\,\int\d^8z\,
\pmtrx{\bJ & J^T } 
\,\gD\,\pmtrx{\bJ^T \\ J}\,\Big\}~.
\labl{FunInteg}
}
To proceed we must invert $\gD^{-1}$. We note that the quadratic
operator $\gD^{-1}$ can be decomposed into a standard free part $P$ and perturbation $L\,$
\equ{
\gD^{-1} = P\inv + L~,\quad P\inv ~=~ \Box\,\pmtrx{0 & P_- \\[1ex] P_+ & 0}~,
\qquad L = \pmtrx{(m + M^2 \bgth^2)\frac{D^2}{-4}  & 0\\[1ex] 0& (m\, + M^2\, \gth^2)\frac{\bD^2}{-4} }~.
}
After a long and tedious computation, the inverse of $\gD^{-1}$ can be cast into the form:
\begin{eqnarray}
\gD = P\,\Big( \Id \,-\,  (L\,P)^2 \Big)\inv\Big(\Id \,-\, L\,P\Big) = 
 \frac{1}{\Box - m^2}\pmtrx{\bC & P_+ \\[1ex] P_- & C}\,+\,\frac{M^2}{(\Box - m^2)^2 - M^4}\,\pmtrx{A_+\,\bC&A_+\\[1ex] A_-& A_-\,C}~,
\label{inverspro}
\end{eqnarray}
where the matrices $A_+$, $A_-$, $C$ and $\bC$ are given by  
\equ{
\arry{cc}{ \dsp 
A_+ ~=~ m\,\get_+\,+\,m\,\bget_+\,+\, \frac{M^2}{\Box - m^2}\Big(\Box\,\get_+\,\bget_+ \,+\,m^2\,\bget_+\,\get_+\Big)~,
\quad & \dsp 
C ~=~ \Big( m\,+\, M^2\, \bgth \Big) 
\frac{D^2}{4\, \Box}~, 
\\[2ex] \dsp 
A_- ~=~ m\,\get_-\,+\,m\,\bget_-\,+\, \frac{M^2}{\Box - m^2}\Big(\Box\,\bget_-\,\get_- \,+\,m^2\,\bget_-\,\get_-\Big)~,
\quad  & \dsp 
\bC ~=~ \Big( m\,+\,M^2\, \gth^2\Big)
 \frac{\bD{}^2}{4\, \Box}~. 
}
}
Notice that $A_- \,=\, A_+^T\,$ and that  $\bC \,=\, C^\dag\,$. Inserting \eqref{inverspro} into \eqref{FunInteg} we obtain the
full propagator with the spurion supersymmetry
breaking~\cite{Helayel-Neto:1984iv}
\begin{eqnarray}
- \frac i 2\int\d^8z\,
\pmtrx{\bJ & J^T } 
\gD
\pmtrx{\bJ^T \\ J}&=& - \frac i 2 \int\d^8z\Big(\bJ_\pm\gD_{\bgF_\pm\gF_\pm}J_\pm
 -  J_\pm\gD_{\gF_\pm\gF_\pm}J_\pm\,+\,\text{h.c.}
  \Big)~.
\label{CPro}
\end{eqnarray}
The notation $\pm$ indicates that we sum over $+$ and $-\,$.  From this expression, we can read off the propagators:
\begin{eqnarray}
\gD_{\bgF_\pm\gF_\pm} &=& \frac{1}{\Box - m^2} \,+\, \frac{M^2}{(\Box - m^2)^2 - M^4}\,\Big\{m\gth^2 + m\bgth^2 +  \frac{M^2}{\Box - m^2}\,\Big(\frac{\bD^2}{-4}\gth^2\,\bgth^2\,\frac{D^2}{-4} + m^2\,\gth^2\,\bgth^2\Big)\Big\}\non\\[2ex]
\gD_{\gF_\pm\gF_\pm} &=& \frac{m}{\Box - m^2}\,\frac{D^2}{-4\,\Box} \,+\, \frac{M^2}{(\Box - m^2)^2 - M^4}\,\Big\{\frac{D^2}{-4}\,\bgth^2 + m^2\frac{D^2}{-4\Box}\,\bgth^2\non\\[2ex]
&&\,+\,  \frac{m\,M^2}{\Box - m^2}\,\bgth^2\,\Big(\frac{D^2}{-4}\gth^2\,+\,\gth^2\frac{D^2}{-4}\Big)\Big\}~.
\label{ChiralProp}
\end{eqnarray}

For the vector multiplets we can perform a very similar analysis to
compute the inverse of the kinetic operator $\gD_V$. Writting the quadratic vector superfield action from \eqref{AcgF2} in terms of projection operator $P_V = 1 - \Id_V$ we find that 
\equ{
S_V ~=~ - 
\,\int\d^8z\, V\,\Big\{\,\gD{}^{-1}_{V}\,-\, J_V\,
\Big\}V~,
\label{AcV3}
}
with the quadratic operator $\gD{}^{-1}_{V}$ and the  identity
matrix  $\Id_V$ given by
\equ{
\gD{}^{-1}_{V} = \Box\,P_V + \Big(\Box\,\Id_V
\,-\, m_\tgg\, \,\Box^{1/2}\,\get_V 
\,-\, m_\tgg\, \,\Box^{1/2}\,\bget_V\,\Big)~,\quad\Id_V ~=~  \get_V\bget_V \,+\, \bget_V\get_V\,.
}
The path integral can be carried out and gives
\equ{
Z_V(J) ~=~  
\int \cD V\, e^{i\, S_V} ~=~ \text{exp}\Big\{\frac i 4 \int \d^8z\, J_V\,\gD_{V}\,J_V\,\Big\}~.
\label{VPro}
}
The vectorfield superpropagator can be read off directly from \eqref{VPro} and reads\footnote{We take this opportunity to point out that the term \(\frac{1}{h\Box - m^2_V}(1 - \Id_v)\) is missing in the expression for the vector propagator eq.(58) in \cite{Nibbelink:2006si}.}
\begin{eqnarray}
\gD_{V} ~&=&~\frac 1{\Box} + \frac{m^2_\tgg}{\Box\,(\Box\,-\,m^2_\tgg)}\,\Id_V\,+\,\frac{m_\tgg}{\Box\,(\Box\,-\,m^2_\tgg)}\,( \,\Box^{1/2}\,\get_V\,+\, \,\Box^{1/2}\,\bget_V)\non\\
[2mm]
~&=&~\frac 1{\Box} + \Bigl\{\frac{m_\tgg}{\Box\,(\Box\,-\,m^2_\tgg)}\Bigl(\frac{m_\tgg}{32\Box}\,D^\ga\bD^2\,\gth^2\,\bgth^2\,D^2\bD^\dga\,\gs^m_{\ga\dga}\,\der_m\,+\,\frac{1}{8}\,D^\ga\bD^2\,\gth^2\,D_\ga\Bigl)\,+\,\text{h.c.}\Bigl\}
\labl{VectorProp}
\end{eqnarray}
In figure \ref{fg:Props} we have collected our graphical
representation for these propagators.
\begin{figure}
\begin{center} 
\mbox{
\tabu{ccccccc}{
$\bgF_\pm$\,\raisebox{-.3ex}{\scalebox{0.5}{\mbox{\input{propbgFgF.pstex_t}}}}
$\gF_\pm$
&$\qquad$ & 
$\gF_\pm$\,\raisebox{-.3ex}{\scalebox{0.5}{\mbox{\input{propgFgF.pstex_t}}}}
$\gF_\pm$
&$\qquad$ & 
$V$\,\raisebox{-.3ex}{\scalebox{0.5}{\mbox{\input{propVV.pstex_t}}}}
$V$
}}
\end{center} 
\captn{This picture gives our drawing conventions for the propagators which we employ throughout this paper. The first two diagrams correspond to the chiral propagators defined in \eqref{ChiralProp}: The first one represents $\gD_{\bgF_\pm\gF_\pm}$ and the second one $\gD_{\gF_\pm\gF_\pm}\,$. The latter refer to the vector propagator given in \eqref{VectorProp}. 
\labl{fg:Props}
}
\end{figure}
\section{Examples}
\label{s:example} 
As a quick application of our result, in this section we compute 
the one-loop chiral multiplet self-energy depicted in figure \ref{fg:selfenergy}, in the presence of soft breaking terms. The classical action and the propagators were given in section \ref{sc:sqed} and \ref{sc:spropa}.  Here we write the vertices, after that we evaluate the Feynman graphs that lead to corrections of the gauge superfield contribution to the two-point $\gF\bgF$-vertices. These contributions come from the following part of \(S_{\rm susy}\) 
\equ{
S_{\rm susy}~\supset~\int \d^8 z\,\Bigl(2\,e\,\bgF\,T\,V\,\gF\,+\,2\,e^2\,\bgF\,V^2\,\gF\Bigl)~.
\labl{Point} 
}
Here we have introduced the charge operator for the electron $T$ 
\equ{
T ~=~ \pmtrx{1 & 0 \\ 0 & -1}~.
}
\subsection{Self Energy due to Selectron Mass}
\label{sc:seselectronMass}
We present examples of calculations of self-energy supergraph depicted in figure \ref{fg:selfenergy}.A. For simplicity, the soft breaking term \(S_{\rm soft}\) is restricted to selectron mass term $M$.
 
The relevant \(2 e \,\bgF\,V\,\gF\) interaction term is given in \eqref{Point}, and 
corresponding propagators are given by $\eqref{ChiralProp}$ and $\eqref{VectorProp}$, but with $m_\tgg = 0$ . To calculate this self energy graph the $\gF$, $\bgF$ and $V$ superfields are replaced by the corresponding functional derivatives with respect to sources that act on the exponential of the propagators \eqref{CPro} and \eqref{VPro}. After functional derivations, the expression for the supergraph figure \ref{fg:selfenergy}.a reads
\equ{
i \gG_\text{A}(m_\tgg = 0) ~=~  \,-\,e^2\,\int (\d^8 z)_{1234}\,\bgF_2\,\,\gF_1\,\frac{D^2_3}{-4}\,\gd_{31}\,(\gD_{\gF\bgF})_3\,\frac{\bD^2_3}{-4}\,\gd_{32}\,\gd_{41}\,(\gD_{V})_4\,\gd_{42}~,
}
where $\gD_{\bgF\gF}\) is given by \eqref{CPro}, and \(\gD_{V} =  \frac{1}{\Box}\). Because (except for $J_V$) all these sources are chiral, the functional differentiation w.r.t. them leads to chiral delta function in superspace: \(\frac{\gd J_{\pm 2}}{\gd J_{\pm 1}} = -\frac{1}{4}\bD^2\gd_{12}\) . Integrating over $ z_3 = (x,\gth)_3$ and $z_4 = (x,\gth)_4$ 
\equ{
\int\d^8 z_3\,F(z_3)\gd_{31} ~=~ F(z_1)~,\qquad \int\d^8 z_4\,F(z_4)\gd_{41} ~=~ F(z_1)~,
}
we are left with
\begin{eqnarray}
i \gG_\text{A}(m_\tgg = 0) &=&  
\,-\,e^2\,\int (\d^8 z)_{12}\,\bgF_2\,\,\gF_1\,\Bigl\{\frac{1}{\Box - m^2}\frac{D^2 \bD{}^2}{16} + \frac{M^2}{(\Box - m^2)^2 - M^4}\Bigl[\Big(m\,\frac{D^2}{-4}\,\gth^2\,\frac{\bD^2}{-4}\ +\text{h.c.}\Bigl)\non\\[2ex] 
&& 
\,+\,\frac{M^2}{\Box - m^2}\,\Bigl(m^2\,\frac{D^2}{-4}\,\gth^2\,\bgth^2\,\frac{\bD^2}{-4} +
M^2\,\frac{D^2}{-4}\frac{\bD^2\,\gth^2\,\bgth^2\,D^2}{16}\,\frac{\bD^2}{-4}\Big)\,\Bigl] 
\Bigl\}_1 \gd_{12} \, \frac 1{\Box_1} \gd_{12}~.
\labl{Onecorrd}
\end{eqnarray}
Using the relations
\equ{
D^2\bD{}^2 D^2 ~=~ 16 \Box\, D^2~,\quad D^2\gth^2 = - 4 + 4\,\gth^\ga D_\ga + \gth^2D^2~,\qquad \bD^2\bgth^2 = - 4 - 4\,\bgth^\dga\bD_\dga + \bgth^2\bD^2~,
}
and the fact that supergraphs are only non--vanishing if
they involve an equal number of super 
covariant derivatives $D$ and $\bD$, and of each of them at least
two
\equ{
\int (\d^8 z)_{12}\,A_2\,\gd_{12}\,[B\,D^2 \bD{}^2]_2\,\gd_{12} ~=~ 16 \int (\d^8 z)_{12}\,A_2\,\gd_{12}\,B_2\,\gd_{12}~,
}
we find that the supergraph, figure \ref{fg:selfenergy}.A, becomes the following scalar integral
\begin{eqnarray}
i \gG_\text{A}(m_\tgg = 0) &=&  
 - \,e^2\, \int (\d^8 z)_{12}\,\bgF_2\,\gF_1\,\Bigl\{\Bigl[\frac{\Box - m^2}{(\Box - m^2)^2 - M^4} + \frac{M^2}{(\Box - m^2)^2 - M^4}\Big(m\gth^2 + m\bgth^2\non\\[2ex]
&& + \frac{M^2}{\Box - m^2}\Bigl[\gth^2\,\bgth^2(\Box + m^2) + 2 i \bgth^\dga\gth^\ga\gs^m_{\dga\ga}\der_m\Bigl]\Big) 
\Bigl]_1\Bigl\} \gd_{12}\frac{1}{\Box_1} \gd_{12}~.
\labl{O}
\end{eqnarray}
As the integral over momentum space is symmetric under \(p \rightarrow - p\), this means that after going to momentum space, the last term of \eqref{O} will not contributes, and hence we will set to zero now. 
The final step in evaluation of this diagram in the coordinate space representation is to make the expression local in the Grassmann variables. By integrating over $\gth_2$
\equ{
\int (\d^8 z)_{12}\,A_2\,\gd_{12}\,B_2\,\gd_{12} ~=~ \int (\d^4 x)_{12}\,\d^4 \gth_1\,A_2\,\gd_{12}^4\,B_2\,\gd_{12}^4~,
}
we find the following expression for \eqref{O}
\begin{eqnarray}
i \gG_\text{A}(m_\tgg = 0) &=&  
- \,e^2\,\int (\d^4 x)_{12}\,\d^4 \gth\,\bgF(x_2,\bgth)\,\gF(x_1,\gth)\,\Bigl[\frac{\Box - m^2}{(\Box - m^2)^2 - M^4} + \frac{M^2}{(\Box - m^2)^2 - M^4}\Big(m\,\gth^2\non\\[2ex]
&& 
+ m\,\bgth^2 +  M^2\Bigl[\gth^2\,\bgth^2\frac{\Box + m^2}{\Box - m^2}
\Bigl]\Big) 
\Bigl]_1 \gd_{12}^4 \, \frac 1{\Box_1} \gd_{12}^4~.
\labl{Onecorrd}
\end{eqnarray}
Since the expression only contains $\gth_1$, it is local in $\gth_1$ and we simply dropped the subscript $"1"$ on $\gth$. In this expression $\gd^4_{21} \,=\, \gd^4(x_2-x_1)\,$ denotes the four dimensional space time delta function, and
the subscript $1$ on the square bracket $[\ldots]$ denote that the corresponding expression is defined in superspace coordinate system $1$. 
The final result in momentum space reads
\begin{eqnarray}
i \gG_\text{A}(m_\tgg = 0) &~=~& 
 \int \frac{\d^4 p}{(4\pi)^4}\int\d^4 \gth\,\Big[\bgF(p,\bgth)\,\gS(p;\gth,\bgth)\,\gF(-p,\gth)\Big]~,
\end{eqnarray}
from which the electron superfields self--energy \(\gS(p;\gth,\bgth)\) can be read off
\begin{eqnarray}
\gS(p;\gth,\bgth)&~=~& - \,e^2\, \int \frac{\d^4 q}{(2\pi)^4}\,\frac{1}{[(q + p/2)^2 + m^2]^2 - M^4}\frac{1}{(q - p/2)^2}\Bigl[(q + p/2)^2 + m^2\non\\
[2ex]
&&
~~~~~- M^2\, m (\gth^2 + \bgth^2)  -  \gth^2\bgth^2\Bigl(M^4 - 2 \frac{m^2\,M^4}{(q + p/2)^2 + m^2}\Bigl)\Bigl]~.
\labl{selfenergy3}
\end{eqnarray}
This integral \eqref{selfenergy3} is divergent and therefore need to be regularized.

The component form of the diagram \eqref{Onecorrd} before eliminating the auxiliary fields \(F_\pm\) and their complex conjugates \(\bF_\pm\) reads
\begin{eqnarray}
i \gG_\text{A}(m_\tgg = 0) &=&  
 - \,e^2\,\int \d^4 x_{12}\Bigl\{\Bigl(\bar{\varphi}_\pm(x_2)\Box\varphi_\pm(x_1) - i \bgps_\pm(x_2)\bgs^m\der_m\gps_\pm^\ga(x_1) + \bF_\pm(x_2) F_\pm(x_1)\Bigl)\times\non\\
[2ex]
&&
\times\Bigl[\frac{\Box - m^2}{(\Box - m^2)^2 - M^4}\Bigl]_1 +  \Bigl(\bF_\pm(x_2)\varphi_\pm(x_1) + \bar{\varphi}_\pm(x_2)F_\pm(x_1)\Bigl)\Bigl[\frac{m\,M^2}{(\Box - m^2)^2 - M^4}\Bigl]_1\non\\
[2ex]
&& + 
\bar{\varphi}_\pm(x_2)\,\varphi_\pm(x_1)
\Bigl[\frac{\Box + m^2}{\Box - m^2}\frac{M^4}{(\Box - m^2)^2 - M^4} 
\Bigl]_1
\,\Bigl\} \gd_{12}^4 \, \frac 1{\Box_1} \gd_{12}^4~.
\labl{Onecorr}
\end{eqnarray}

As it stands this expression \eqref{Onecorr} is logarithmically divergent and
requires regularization. We have chosen to use dimensional reduction 
\cite{Siegel:1979wq,Capper:1979ns}. In appendix \ref{sc:basicintegrals} we have collected the one loop integrals calculated in 
this scheme. Using the standard scalar integrals $J_2\,$ and $J_3\,$ defined in appendix
\ref{sc:basicintegrals} (see \ref{IJ}), we obtain 
\begin{eqnarray}
\gG_\text{A}(m_\tgg = 0) &=& 
 -  \frac{1}{2}e^2\int \frac{\d^4 p}{(4\pi)^4}\Bigl\{\Bigl(\bF_\pm(p) F_\pm(-p) - \bar{\varphi}_\pm(p)p^2\varphi_\pm(-p) - i \bgPs_D(p) p\slashed\gPs_D(-p)\Bigl)
 \int\limits_0^1\!\d\,x\Bigl[J_2(\tm_-^2)\non\\
 [2mm]
&& 
+ J_2(\tm_+^2)\Bigl]
+ \Bigl(\bF_\pm(p)\varphi_\pm(-p) + \bar{\varphi}_\pm(p)F_\pm(-p) + M^2\bar{\varphi}_\pm(p)\varphi_\pm(-p)\Bigl)
 \int\limits_0^1\!\d\,x\Bigl[J_2(\tm_-^2)\non\\
 [2mm]
&& - J_2(\tm_+^2)\Bigl]
+ 4\,m^2\,M^2\,\bar{\varphi}_\pm(p)\varphi_\pm(-p)\,\int\limits_0^1\!\d\,x\,\int\limits_0^{1 - x}\!\d\,y\Bigl[J_3(\tM_-^2) - J_3(\tM_+^2)\Bigl]
\Bigl\}~.
\label{sescalar}
\end{eqnarray} 
In this expression we have combined the two Weyl spinors $\gps_\pm$ into a charged Dirac spinor \eqref{Dirac}, the electron. The masses $\tm_\pm$ and $\tM_\pm$ are defined by
\equ{
\tm_\pm^2 \,=\, x(1 - x)p^2 + \,x\, m^2_\pm~,\quad \tM^2_\pm \,=\, m^2_\pm y + m^2(1 - x - y) + x(1 - x)p^2~,
}
with  $ m_\pm^2 \,=\, m^2\,\pm \, M^2\, $.  
From the the one-loop result given in \eqref{sescalar} the wave function renormalization for the electron and its superpartner the selectron can be obtained from the divergent integral:
\begin{eqnarray} 
\frac{e^2}{2}\int\limits_0^1\!\d\,x\Bigl[J_2(\tm_-^2) + J_2(\tm_+^2)\Bigl]
&~=~& \frac{e^2}{16\pi^2}\frac{1}{\bge} + \frac{e^2}{16\pi^2}\Bigl[- 2 + \ln \frac{m^2_\pm}{\gm^2}
+ \frac{p^2 + m^2_\pm}{p^2}  \ln \frac{p^2 +m^2_\pm}{m^2_\pm}\Bigl]~.
\end{eqnarray} 
where \(\frac{1}{\bge} = 1\ge - \gg_E + \ln(4\pi)\). From this expression, we can read off the wave function renormalization for the electron  \cite{DeWit:1986it} and selectron
\begin{eqnarray} 
Z_1 ~=~ Z_{\bar{\varphi}_\pm\varphi_\pm} ~=~ Z_{\bgPs_D\gPs_D} ~=~  \gm^{-2\ge} \Bigl(1 - \frac{e^2}{16\pi^2}\frac 1\bge\Bigl)~.
\labl{wavefun}
\end{eqnarray} 
As for the divergences arising from the integrals \(\int\limits_0^1\!\d\,x[J_2(\tm_-^2) - J_2(\tm_+^2)]\) cancel each other, and hence the second and third line of \eqref{sescalar} are finite renormalizations for the selectron soft mass $M$. 

This computation confirm that the divergent part of the wave function renormalization is independent of the soft breaking, and is the same for selectron and the electron.  A consistency check on this result is obtained when one considers the well-known one beta functions \cite{Inoue:1982pi,Inoue:1983pp,Derendinger:1983bz,Gato:1984ya,Falck:1985aa} (for results including two loop beta functions see
\cite{Martin:1993zk,Yamada:1994id})  for the parameters $M$ and $m$, if we restrict to renormalizable models. In our previous paper \cite{Nibbelink:2006si}, we obtain these beta functions by computing the renormalization of the parameters $M$ and $m$, and found exact agreement.  The absence of one-loop corrections to the electron mass $m$ is a result of the well known non--renormalization theorem for the superpotential \(\int \d^6 z\,( m\,\gF_+ \gF_- + \text{h.c.})\) \cite{Grisaru:1979wc} (see also \cite{Weinberg:1998uv,Poppitz:1996na,Seiberg:1993vc}). Due to the non--renormalization, one also obtained the electron mass $m$ is renormalized via the wave function renormalization $Z_1$ like 
\equ{
m ~=~ Z_1^{-1}m_R~,
}
where $m_R$ is the renormalized mass (For detais see the texbook \cite{Buchbinder:1998qv}).

\begin{figure}
\begin{center} 
\mbox{
\tabu{ccccccc}{
\raisebox{-3.3ex}{\scalebox{0.6}{\mbox{\input{chiral_self.pstex_t}}}}
&\qquad \qquad &
\raisebox{-1.3ex}{\scalebox{0.6}{\mbox{\input{chiral_tad.pstex_t}}}}
\\[2ex]
A&& B 
\\[3ex] 
}}
\end{center} 
\captn{One-loop diagrams contributing to electron and selectron self energies.
\labl{fg:selfenergy}
}
\end{figure}
\subsection{Self Energy due to Photino Mass}
\label{sc:sephotinoMass}
We compute the one-loop self energy corrections due to the soft photino mass insertions depicted \ref{fg:selfenergy}. We first consider the supergraph figure \ref{fg:selfenergy}.A.  After functional derivations, the expression for the supergraph figure reads
\equ{
i \gG_\text{A}(M = 0) ~=~  \,-\,e^2\,\int (\d^8 z)_{12}\,\bgF_2\,\gF_1\Bigl[\gD_{\gF\bgF}\frac{D^2\bD^2}{16}\Bigl]_1\gd_{12}\,(\gD_{V})_1\,\gd_{12}~,
}
where \(\gD_{\bgF\gF}\) and \(\gD_{V}\) are respectively given by \eqref{CPro} and  \eqref{VectorProp}, but with selectron mass \(M = 0\). To reduce this integral to a scalar integral we partially integrate the \(\bD^2D^2\) that 
acts on the first \(\gd_{21}\), to obtain
\begin{eqnarray}
i \gG_\text{A}(M = 0) &=& - \frac{e^2}{16}\,\int\d^8 z_{12}\bgF_2\Bigl\{\gF_1(\gD_{\bgF\gF})_1\gd_{12}[\bD^2 D^2]_1 + (\bD^2 D^2\gF)_1(\gD_{\bgF\gF})_1\gd_{12} \non\\
[2ex]
& & + 2(\bD_\dga D^2\gF)_1(\gD_{\bgF\gF})_1\gd_{12}\bD^\dga_1
 - 2 (D_\ga\gF)_1(\gD_{\bgF\gF})_1\,\gd_{12}[\bD^2D^\ga]_1
 \non\\
[2ex]
& & - 4 (\bD_\dga D_\ga\gF)_1(\gD_{\bgF\gF})_1\gd_{12}[\bD^\dga D^\ga]_1
+ (D^2\gF)_1(\gD_{\bgF\gF})_1\gd_{12}\bD^2_1\Bigl\}(\gD_{V})_1\gd_{12}~.
\label{photino}
\end{eqnarray}
Observe here that after partial integrating the \(\bD^2D^2\) that 
acts on the first \(\gd_{21}\), the spurion operators $\Id_V$ and $\get_V\,$ (which are hidden in vector superpropagator $\gD_{V}$) find themselves
surrounded by  two superspace delta functions $\gd_{21}$.  By integrating over the full double superspace, we obtained the identities given in appendix \ref{sc:idendities}. Using these identities (\ref{a1}--\ref{a7}), the last term of \eqref{photino} vanishes, 
and we find that remaining terms becomes the following scalar integral
\begin{eqnarray}
i \gG_\text{A}(M = 0) &=& -\,e^2\,\int\d^4 x_{12}\d^4\gth\,\bgF_2\Bigl\{\gF_1\frac{1}{(\Box - m^2)_1}\gd^4_{12}\frac{1}{\Box_1} -  m^2_\tgg\,D_\gb\gF\,\gth^\gb\frac{1}{(\Box - m^2)_1}\gd^4_{12}\Bigl[\frac{1}{\Box(\Box - m^2_\tgg)}\Bigl]_1\non\\
[2ex]
&&
 \,+ \,m_\tgg\,\bD_\dgb D^2\gF_1\Bigl(\frac{1}{4}\bgth^\dgb - \frac{1}{2}\gth^2\bgth^\dgb\,m_\tgg\Bigl)\frac{1}{(\Box - m^2)_1}\gd^4_{12}\Bigl[\frac{1}{\Box(\Box - m^2_\tgg)}\Bigl]_1\non\\
[2ex]
&&
\,-\, m^2_\tgg\,(\bD_\dgb D_\gb\gF)_1\frac{1}{(\Box - m^2)_1}\gd^4_{12}\Bigl[\bgth^\dgb\gth^\gb\frac{1}{\Box(\Box - m^2_\tgg)}  + \bgth^\dga\gth^\gg\,\ge^{\gb\ga}\,\ge^{\dgb\dgg}\gs^m_{\dga\ga}\gs^n_{\dgg\gg}
\frac{\der_m\der_n}{\Box^2(\Box - m^2_\tgg)}
 \Bigl]_1
\non\\
[2ex]
&&
 \,+ \,2 \,m_\tgg\,(\Box\gF)_1\Bigl(\gth^2 + \bgth^2 -  2 \bgth^2\gth^2\,m_\tgg
 \Bigl)\frac{1}{(\Box - m^2)_1}\gd^4_{12}\Bigl[\frac{1}{\Box(\Box - m^2_\tgg)}
 \Bigl]_1\Bigl\}\gd^4_{12}~,
\labl{photino1}
\end{eqnarray}
where we have discarded the terms containing $\der_m$ because they are antisymmetric under $\der_m\rightarrow \der_m\,$.

Next we turn to supergraph figure \ref{fg:selfenergy}.B\,. The two-point vertex \(2 e^2 \bgF\,V^2\,\gF\) that give rise to that supergraph is obtained from \eqref{Point}. Using standard supergraphs techniques we find that the supergraph figure \ref{fg:selfenergy}.B, becomes the following scalar integral  
\equ{
i \gG_\text{B} ~=~  
\,e^2\,\int (\d^8 z)_{12}\,\bgF_1\,\gF_1\,\gd_{21}\,(\gD_{V})_2\,\gd_{21}~,
\labl{supgraphB}
}
with the vector superfield propagator given in \eqref{VectorProp}. Upon using \ref{a1} in appendix \ref{sc:idendities} the supergraph \eqref{supgraphB} becomes the following scalar integral
\begin{eqnarray}
i \gG_\text{B} &=&  
\,e^2\,\int \d^4 x_{12}\,\d^4\gth\,\bgF_1\,\gF_1\,\gd^4_{21}\,\Bigl[2\,\frac{m_\tgg}{\Box(\Box - m_\tgg^2)}\Bigl(\,\gth^2\,+\,\bgth^2\Bigl)\,-\,4\,\frac{m_\tgg^2}{\Box(\Box - m_\tgg^2)}\gth^2\,\bgth^2\Big)\Bigl]_2 \gd^4_{21}~.
\labl{supgraphB1}
\end{eqnarray}

By combining these results \eqref{photino1} and \eqref{supgraphB1}, and computing the component action, we find 
\begin{eqnarray}
i \gG_{\text{full}}  &~=~& i \gG_\text{A}(M = 0) + i \gG_\text{B}\non\\
[2mm]
 &~=~& - \,e^2\, \int\d^4 x_{12}\Bigl\{\Bigl(\bF_\pm(x_2)F_\pm(x_1) + \bar{\varphi}_{\pm}(x_2)\Box\varphi_{\pm}(x_1) - i\gps_{\pm}(x_1)\gs^m\der_m\bgps_{\pm}(x_2)\Bigl)\times\non\\
[2ex]
&&
\times
\frac{1}{(\Box - m^2)_1}\gd^4_{12}\frac{1}{\Box_1}\gd^4_{12}
+ \,m^2_\tgg\,\Bigl(2 \,\bF_\pm(x_2)F_{\pm}(x_1) + 2 \,\bar{\varphi}_{\pm}(x_2)\Box\varphi_{\pm}(x_1)\non\\
[2ex]
&& + i\,5\,\bgps_{\pm}(x_2)\bgs^m\der_m\gps_{\pm}(x_1)\Bigl)
\frac{1}{(\Box - m^2)_1}\gd^4_{12}\Bigl[\frac{1}{\Box(\Box - m^2_\tgg)}\Bigl]_1 \gd^4_{12}
-\,m^2_\tgg\,\Bigl(2 \bar{\varphi}_{\pm}(x_2)\der^m\der^n\varphi_{\pm}(x_1)
\non\\
[2ex]
&&  + i\,2\,\bgps_{\pm}(x_2)\bgs^m\der^n\gps_{\pm}(x_1) + n \leftrightarrow m\Bigl)
\frac{1}{(\Box - m^2)_1}\gd^4_{12}\Bigl[\frac{\der_m\der_n}{\Box^2(\Box - m^2_\tgg)}\Bigl]_1
\gd^4_{12}\\
[2mm]
&&
- \,2m_\tgg\,\Bigl(\bF_\pm(x_1)\varphi_\pm(x_1) + \bar{\varphi}_\pm(x_1)F_\pm(x_1)
 -  2\,m_\tgg\,\bar{\varphi}_\pm(x_1)\,\varphi_\pm(x_1)\Bigl)\gd^4_{12}\Bigl[\frac{1}{\Box(\Box - m^2_\tgg)}\Bigl]_1 \gd^4_{12}
\Bigl\}\non~.
\labl{comPaction}
\end{eqnarray}

This expression can be evaluated further using the same Fourier
transforms and Wick rotations as employed for diagram $\gG_A(m_\tgg = 0)$, and find
\begin{eqnarray}
\gG_\text{full} &~=~& 
- \,e^2\,\int \frac{\d^4 p}{(4\pi)^4}\,\Bigl\{\Bigl(\bF_\pm(p)F_{\pm}(-p) - \bar{\varphi}_\pm(p)p^2\varphi_\pm(-p)  + i \bgPs_D(p)p\slashed\gPs_D(-p)\Bigl)L(p^2, m^2) 
\non\\
[2ex]
&&
+ \,\tm^2_\gg
\Bigl(2\,\bF_\pm(p)F_{\pm}(-p) - 2\,\bar{\varphi}_\pm(p)p^2\varphi_\pm(-p) - i 5\,\bgPs_D(p)p\slashed\gPs_D(-p)\Bigl)\,L(p^2, \tM^2)\non\\
[2ex]
&&
+\,m^2_\tgg\,\Bigl(2 \bar{\varphi}_{\pm}(p)p^mp^n\varphi_{\pm}(-p)  + i \,\bgps_D(p)\gg^mp^n\gps_D(-p) + n \leftrightarrow m\Bigl)L_{nm}({p^2, \cal{\tM}}^2)\non\\
[2ex]
&&
- \,2m_\tgg\,\Bigl(\bF_\pm(p)\varphi_\pm(-p) + \bar{\varphi}_\pm(p)F_\pm(-p) -  2\,m_\tgg\,\bar{\varphi}_\pm(p)\,\varphi_\pm(-p)\Bigl)\,I(m^2_\tgg)
\labl{fullselfen}
\Bigl\}~.
\end{eqnarray} 
The integrals $I(m^2_\tgg)\,$, and $L(p^2, m^2)\,$ are divergent, whereas $L(p^2,\tM^2)\,$ and $L_{nm}({p^2, \cal{\tM}}^2)\,$ are finite. We have evaluated these integrals in appendix \ref{sc:basicintegrals}, (see \eqref{IJ}, \eqref{B22}, 
\eqref{B23} and \eqref{B24}) respectively, with
\begin{eqnarray}
\tm^2 &~=~& x(1 - x)p^2 + x m^2~,\qquad \tM^2 ~=~ y(1 - y)p^2 + m^2_\tgg(1 - x - y)  + m^2 y~,\\
[2mm]
{\cal{\tM}}^2  &~=~& m^2 y + m^2_\tgg(1 - x - y)~.
\end{eqnarray}
From this result we can determine the renormalized quantities and a number of additional finite terms that are second order in the coupling constant $e\,$.  The wave function renormalization $Z_1$ is obtained from the integral $L(p^2, m^2)$:
\begin{eqnarray}
L(p^2, m^2) = \gm^{-2\ge}\,\int\limits_0^1\!\d\,x\,J_2(m^2) ~=~ \frac {\gm^{-2\ge}}{16 \pi^2}\Bigl[  
\frac 1\bge + 2 - \ln \frac{m^2}{\gm^2} -  \frac{p^2 + m^2}{p^2}\ln \frac{p^2 + m^2}{m^2}
\Bigl]~,
\end{eqnarray}
from which we read off the wave function renormalization $Z_1$
\begin{eqnarray} 
Z_1 ~=~  \gm^{-2\ge} \Bigl(1 - \frac{e^2}{16\pi^2}\frac 1\bge\Bigl)~,
\labl{wavefun1}
\end{eqnarray} 
that is consistent with \eqref{wavefun}. Because of the non-renormalization theorem, we also find that the electron mass $m$ did not received quantum correction except for the wave function renormalization: \(m ~=~ Z_1^{-1}m_R\). The second and the third line of \eqref{fullselfen} are additional finite renormalization for the photino mass.
\section{Counterterms}
\labl{sc:counterterms}
As a first step towards the renormalization we needs to introduce counterterms. From the first example in subsection \ref{sc:seselectronMass}, the infinities arises in the first term of \eqref{Onecorrd}
\begin{eqnarray}
i \gG_\text{A}(m_\tgg = 0) &~=~&  
- \,e^2\,\int (\d^4 x)_{12}\,\d^4 \gth\,\bgF(x_2,\bgth)\,\gF(x_1,\gth)\,\Bigl(\frac{\Box - m^2}{(\Box - m^2)^2 - M^4}\Bigl)\gd^4_{12}\frac{1}{\Box_1}\gd^4_{12}~,
\end{eqnarray}
which after regularization becomes 
\begin{eqnarray}
\gG_\text{A}(m_\tgg = 0) &~=~& -\,\frac{e^2}{2}\int \frac{\d^4 p}{(4\pi)^4}\int\limits_0^1\!\d\,x\Bigl[J_2(\tm_-^2) + J_2(\tm_+^2)\Bigl]\,\int\,\d^4\gth\,\,\bgF_\pm(p,\bgth)\gF_\pm(-p,\gth)~. 
\end{eqnarray}
The divergent part of the above expression reads
\begin{eqnarray}
\gG^{\text{div}}_\text{A}(m_\tgg = 0) &~=~& -\,\frac{e^2\gm^{-2\ge}}{16\pi^2}\frac{1}{\bge} \,\int \frac{\d^4 p}{(4\pi)^4}\,\int\,\d^4\gth\,\,\bgF_\pm(p,\bgth)\gF_\pm(-p,\gth)~.
\labl{div}
\end{eqnarray}
In the second example in subsection \ref{sc:sephotinoMass} we find that the divergent part take the same form as \eqref{div}
\equ{
\gG^{\text{div}}_\text{A}(m_\tgg = 0) ~=~ \gG^{\text{div}}_\text{A}(M = 0)~.
}
Finally, from the diagram $\gG_B$ give rise to infinite terms that involve spurion superfields $\gth^2$ and $\bgth^2$: 
\begin{eqnarray}
\gG_B^{\text{div}} &=&  2 m_\tgg\frac{e^2\gm^{-2\ge}}{16\pi^2}\frac{1}{\bge}\,\int \frac{\d^4 p}{(4\pi)^4}\,\int\,\d^4\gth\,\Bigl\{\Bigl(\gth^2 + \bgth^2 - 2\,m_\tgg\,\gth^2\bgth^2\Bigl)\bgF_\pm(p,\bgth)\gF_\pm(-p,\gth)\Bigl\}~.
\labl{div1}
\end{eqnarray}
To cancel these infinities \eqref{div} and \eqref{div1} we have to introduce two one-loop counter terms
\equ{
\gD S ~=~ \gD S_1 + \gD S_2~.
}
The first term cancels the divergent part \eqref{div} of the electron self-enegy diagram $\gG_A$ and takes the form
\begin{eqnarray}
\gD S_1 ~=~ \gD Z_1\,\dsp \int \d^8 z\,\Bigl( 
\bgF_+\,\gF_+ + \bgF_-\,\gF_-\Bigl)~,
\label{count}
\end{eqnarray}
where \(\gD Z_1 ~=~  Z_1 - 1\). The second counterterm is equal to
\begin{eqnarray}
\gD S_2 &~=~&  2\,m_\tgg\,\gD Z_1\,\dsp \int \d^8 z\,\Bigl(- \gth^2 - \bgth^2 + 2 m_\tgg\gth^2\bgth^2\Bigl)\,\Bigl( 
\bgF_+\,\gF_+ + \bgF_-\,\gF_-\Bigl)~.
\labl{divergent}
\end{eqnarray}
Addition of the this term cancels the infinite part  \eqref{div1} of diagram $\gG_B$. 

The fact that loop correction generates divergencies that contains spurion superfields is not surprising. This issue has been discussed before in the literature \cite{Girardello:1981wz,Chang:1985qd,Yamada:1994id}. Following Yamada \cite{Yamada:1994id} these divergencies can be removed by spurion dependent transformations
\equ{
\gF^\prime_\pm ~=~ \gF_\pm - 2  m_\tgg\,\gth^2\,\gF_\pm~,\quad \bgF^\prime_\pm ~=~ \bgF_\pm - 2  m_\tgg\,\bgth^2\,\bgF_\pm~.
}
In summary, we have shown from these examples that there is only one counterterm needed and it is given by
\begin{eqnarray}
\gD S^\prime ~=~ \gD Z_1\,\dsp \int \d^8 z\,\Bigl( 
\bgF^\prime_+\,\gF^\prime_+ + \bgF^\prime_-\,\gF^\prime_-\Bigl)~.
\label{count}
\end{eqnarray}
\section{Summary} 
\label{sc:sum} 
Supergraphs is a powerful, elegant and efficient tool for computations even when supersymmetry is softly broken. In these cases supersymmetry breaking are represented by spurions.

In this paper we have  represented supersymmetry breaking by spurion superfields, and  worked out explicit expressions for the propagators. By using them we calculated exact expressions for the one-loop diagrams that contributes to electron and selectron self energies. In particular, we calculate the one-loop gauge superfield contribution to the two-point function from which one can extract the wave function renormalization for the electron and its superpartner the selectron. Our computation confirm that the divergent part of the wave function renormalization is independent of the soft breaking, and is the same for selectron and the electron. Furthermore, we found there are divergencies that involve spurion superfields which can be removed by Yamada spurion dependent transformations.
\section*{Acknowledgments}
We would like to thank Stefan Groot Nibbelink for numerous discussions, encouragement, reading the manuscript. We would also like to thank Jan-Willem van Holten for suggesting many useful remarks.
\appendix 
\def\theequation{\thesection.\arabic{equation}} 
\setcounter{equation}{0}
\section{Identities}
\labl{sc:idendities}
In this appendix we give the relations which we have used to reduce the supergraph integral \eqref{photino} to scalar integral \eqref{photino1}. By integrating over the full double superspace, we find the following relations
\begin{eqnarray}
\int\d^8z_{12}\, \gd_{12}\, [\gD_{V}]_1\, \gd_{12}
~&=&~ 2  \int\d^4\,x_{12} \d^4\gth\, \gd^4_{12}\,\Bigl[\frac{m_\tgg}{\Box\,(\Box\,-\,m^2_\tgg)}\Bigl(\gth^2 + \bgth^2 - 2\,\gth^2\bgth^2\,m_\tgg\non\\
[2ex]
&&
~~~~~~~~~~~~~~~~~~~~~~~~~~~ + \,i\, \bgth^\dga\,\gth^\ga\,\gs^m_{\dga\ga}\frac{\der_m}{\Box}\,m_\tgg\Bigl)\Bigl]_1\gd^4_{21}
\labl{a1}~,
\\[2ex]
\int\d^8z_{12}\, \gd_{12}\, [\bD^2D^2\gD_{V}]_1\, \gd_{12}
~&=&~ 16 \int\d^4\,x_{12} \d^4\gth\, \gd^4_{12}\,\frac{1}{\Box}_1\, \gd^4_{21}
\labl{a2}~,\\[2ex]
\int\d^8z_{12}\, \gd_{12}\, [\bD^\dgb\gD_{V}]_1\, \gd_{12}
~&=&~ - \,2 \int\d^4\,x_{12} \d^4\gth\,\gd^4_{12}\,\Bigl[\Bigl(2\,\gth^2\bgth^\dgb - \,i\,\ge^{\dgb\dga}\,\gth^\ga\gs^m_{\dga\ga}\frac{\der_m}{\Box}\Bigl)\frac{m^2_\tgg}{\Box\,(\Box\,-\,m^2_\tgg)}\non\\
[2ex]
&&~~~~~~~~~~~~~~~~~~~~~~~~~~~~~ - \frac{m_\tgg}{\Box\,(\Box\,-\,m^2_\tgg)}\bgth^\dgb\Bigl]_1\gd_{12}
\labl{a3}~,\\
[2ex]
\int\d^8z_{12}\gd_{12}[\bD^2 D^\gb\gD_{V}]_1\gd_{12}~&=& ~8 \int\d^4\,x_{12} \d^4\gth\,\gd^4_{12}\Bigl[\Bigl(
\gth^\gb \,m_\tgg 
- i\,\bgth_\dgb\bgs^{m\dgb\gb}\der_m\Bigl)\frac{m_\tgg}{\Box\,(\Box\,-\,m^2_\tgg)}\Bigl]_1\gd_{12}
\labl{a4}~,\\
[2ex]
\int\d^8z_{12}\, \gd_{12}\, [\bD^2\gD_{V}]_1\, \gd_{12}
~&=&~ 0
\labl{a5}~,\\
[2ex]
\int\d^8z_{12}\, \gd_{12}\, [\bD^\dgb D^\gb\gD_{V}]_1\, \gd_{12}~&=&~ - 2 \int\d^4\,x_{12} \d^4\gth\,\gd^4_{12}\,\Bigl[\Bigl((i \bgth^2\gth^2\bgs^{m\dgb\gb}\der_m - 2 \bgth^\dgb\gth^\gb)m_\tgg  +  i\,(\bgth^2 + \gth^2)\bgs^{m\dgb\gb}\der_m\non\\
[2ex]
&&
~~~~~~~~~ - 2\,\bgth^\dga\gth^\gg\,\ge^{\gb\ga}\,\ge^{\dgb\dgg}\gs^m_{\dga\ga}\gs^n_{\dgg\gg}\frac{\der_m\der_n}{\Box}m_\tgg\Bigl)\frac{m_\tgg}{\Box\,(\Box\,-\,m^2_\tgg)}\Bigl]_1\gd_{12}
\labl{a6}~,\\
[2ex]
\int\d^8z_{12}\, \gd_{12}\, [D^\gb\gD_{V}]_1\, \gd_{12}~&=&~  2 \int\d^4\,x_{12} \d^4\gth\,\gd^4_{12}\,\Bigl[\Bigl(2\,\bgth^2\gth^\gb - \,i\,\ge^{\gb\ga}\,\bgth^\dga\gs^m_{\dga\ga}\frac{\der_m}{\Box}\Bigl)\frac{m^2_\tgg}{\Box\,(\Box\,-\,m^2_\tgg)}\non\\
[2ex]
&&~~~~~~~~~~~~~~~~~~~~~~~~~~~~~~~~~~~~ + \frac{m_\tgg}{\Box\,(\Box\,-\,m^2_\tgg)}\gth^\gb\Bigl]_1\gd_{12}
\labl{a7}~.
\end{eqnarray}
\section{One Loop Scalar Integrals}
\labl{sc:basicintegrals}
This appendix is devoted to the evaluation of the basic one loop scalar
integrals, which arise in the main text of this paper. We compute these
scalar integrals in the \MSbar\ scheme: We evaluate the integrals in  
$D ~=~ 4 - 2 \ge$ dimensions, and we introduce the renormalization
scale $\gm$ such that all $D$ dimensional integrals have the same mass
dimensions as their divergent four dimensional counter parts.

The three basic type one loop integrals are given by 
\begin{eqnarray}
J_\ga(M^2) &~=~& \int \frac{\d^D p}{(2\pi)^D \gm^{D-4}} \,  \frac{1}{(p^2 + M^2)^\ga} 
~=~ \frac {1}{16 \pi^2} \, \frac{1}{(M^2)^{\ga - 2}}
\Big( 4\gp \frac {\gm^2}{M^2}  \Big)^{2 - \frac D2}
\, \frac{\gG(\ga - \frac D2)}{\gG(\ga)}~,\\
[2mm]
T_{\ga mn}(M^2) &~=~& \int \frac{\d^D p}{(2\pi)^D \gm^{D-4}} \,  \frac{p_m\,p_n}{(p^2 + M^2)^\ga}  
~=~ \frac{1}{D}\eta_{mn}\Bigl(J_{\ga - 1}(M^2) - M^2 J_\ga(M^2)\Bigl)~,\\
[2mm]
\label{J}
I(M^2) &~=~& \int \frac{\d^D p}{(2\pi)^D \gm^{D-4}} \, 
\frac 1{p^{2}}\, \frac{1}{p^2 + M^2} 
~=~ \frac {1}{16 \pi^2} \, 
\Big( 4\gp \frac {\gm^2}{M^2}  \Big)^{2- \frac D2}
\, \frac 1{\gG(\frac{D}{2})}\, \frac {\gp}{\sin \gp(\frac{D}{2}-1)}
~, 
\label{I}
\end{eqnarray}
for $\ga = 2, 3, 4\,$. In the applications in the main text we need to expand
this to the zeroth order in $\ge$ including the pole $1/\ge\,$:
\begin{eqnarray}
J_2(M^2) &~=~& \frac {1}{16 \pi^2}  
\Big[  
\frac 1\ge  - \ln \frac{M^2}{\bgm^2} 
\Big]~, 
~~~~~~~\qquad J_3(M^2) ~=~\frac {1}{32 \pi^2}\frac{1}{M^2}~,
\labl{J23}
\\
[2mm]
I(M^2) &~=~& \frac {1}{16 \pi^2} 
\Big[  
\frac 1\ge + 1 - \ln \frac{M^2}{\bgm^2}\Bigl]~,~\qquad J_4(M^2) ~=~ \frac {1}{96 \pi^2}\frac{1}{M^4}~. 
\labl{IJ}
\end{eqnarray}
Here we have introduced the \MSbar\ scale $\bgm^2 \,=\, 4\gp e^{-\gg_E} \gm^2$ with Euler constant $\gg_E$. 

The second three integrals we encounter in this work are
\begin{eqnarray}
L(p^2, M^2) &~=~& \, \int \frac{\d^D q}{(2\pi)^D \gm^{D-4}} \,\frac{1}{(q +p/2)^2 + M^2}\frac{1}{(q - p/2)^2}~,\\
[2mm]
\label{Ln}
L(p^2, M_1^2, M_2^2) &~=~& \, \int \frac{\d^D q}{(2\pi)^D \gm^{D-4}} \,\frac{1}{(q + p/2)^2 + M_1^2}\frac{1}{(q - p/2)^2 + M_2^2}\frac{1}{(q - p/2)^2}~,\\
[2mm]
\label{Ln}
L_{mn}(p^2, M_1^2, M_2^2) &~=~& \, \int \frac{\d^D q}{(2\pi)^D \gm^{D-4}} \,\frac{q_m q_n}{[(q + p/2)^2 + M_1^2][(q - p/2)^2 + M_2^2][(q - p/2)]^4}~.
\label{Lnn}
\end{eqnarray}
Using Feynman parametrizations
\begin{eqnarray}
\frac{1}{AB} &~=~&
\,\int\limits_0^1\!\d\,x\,\frac{1}{[xA + (1 - x)B]^2}~,
\\
[2mm]
\frac{1}{ABC} &~=~&
2 \,\int\limits_0^1\!\d\,x\,\,\int\limits_0^{1- x}\!\d\,y\frac{1}{[xA + yB + (1 - x - y)C]^3}\\
[2mm]
 \frac{1}{A^2BC}&~=~& 6 \,\int\limits_0^1\!\d\,x\,\int\limits_0^{1 - x}\!\d\,y\frac{x}{[xA + yB + (1 - x - y)C]^4}~,
\end{eqnarray}
we rewrite the $L(p^2, M^2)\,$ $L(p^2, M^2_1, M^2_2)$ and $S_{mn}(M^2_1, M^2_2)$ in a more convenient form, namely
\begin{eqnarray}
L(\tm^2) &~=~& \, \frac{1}{2}\,\int\limits_0^1\!\d\,x\,\int \frac{\d^D \tq}{(2\pi)^D \gm^{D-4}}\,\frac{1}{(\tq^2 + \tm^2)^2}\\
[2mm]
L(\tM^2) &~=~& 2 \,\int\limits_0^1\!\d\,x\,\,\int\limits_0^{1 - x}\!\d\,y\,\int \frac{\d^D q^\prime}{(2\pi)^D \gm^{D-4}}\,\frac{1}{(q^{\prime 2} + \tM^2)^3}\\
[2mm]
L_{mn}(p, {\cal{\tM}}^2) &~=~& 6 \,\int\limits_0^1\!\d\,x\,\,\int\limits_0^{1 - x}\!\d\,y\,\int \frac{\d^D q^\prime}{(2\pi)^D \gm^{D-4}}\,\Bigl[x\,\frac{q^{\prime}_mq^{\prime }_n}{(q^{\prime 2} + {\cal{\tM}}^2)^4} + x\,y^2 \frac{p_m p_n}{(q^{\prime 2} + {\cal{\tM}}^2)^4}\Bigl]~.
\end{eqnarray}
where
\begin{eqnarray}
\tm^2 &~=~&x(1 - x)p^2 + xM^2~,\qquad \tM^2 ~=~ y(1 - y)p^2 + M_1^2 y + M^2_2(1 - x - y)\non\\
[2mm]
{\cal{\tM}}^2  &~=~& y M_1^2  + (1 - x - y)M^2_2~.
\end{eqnarray}
In obtaining these formula we have shifted the integration variables $ \tq \,=\, q + (x - \frac{1}{2})p^2\, $ and $ q^\prime \,=\, q - \frac{1}{2}(1 - 2 v )p\, $. It is not difficult to confirm that these integrals can be written in terms of the simple integrals $J_2$ and $J_3$ as
\begin{eqnarray}
L(p^2, \tm^2)  &~=~& \,  \,\int\limits_0^1\!\d\,x\,J_2(\tm^2) ~=~ \frac {1}{16 \pi^2}\Bigl[  
\frac 1\ge + 2 - \ln \frac{M^2}{\bgm^2} -  \frac{p^2 + M^2}{p^2}\ln \frac{p^2 + M^2}{M^2}
\Bigl]\non\\
[2mm]
&~=~& \frac {1}{16 \pi^2}\Bigl[  
\frac 1\ge - \gg + \ln(4\pi) + 2 - \ln \frac{M^2}{\gm^2} -  \frac{p^2 + M^2}{p^2}\ln \frac{p^2 + M^2}{M^2}
\Bigl]
\\
[2ex]
\labl{B22}
L(p^2,\tM^2) &~=~&2 \,\int\limits_0^1\!\d\,x\,\,\int\limits_0^{1 - x}\!\d\,y\,J_3(\tM^2) ~=~ \frac {1}{16 \pi^2} \,\int\limits_0^1\!\d\,x\,\,\int\limits_0^{1 - x}\!\d\,y\,\frac {1}{\tM^2}
\labl{B23}\\
[2ex]
L_{ mn}({p^2, \cal{\tM}}^2) &~=~& 6 \,\int\limits_0^1\!\d\,x\,\,\int\limits_0^{1 - x}\!\d\,y\,\Bigl[x\Bigl\{T_{4mn}({\cal{\tM}}^2) + y^2 p_m p_n J_4({\cal{\tM}}^2)\Bigl\}\Bigl]\non\\
&~=~& \frac 1{2 \pi^2} \,\int\limits_0^1\!\d\,x\,\,\int\limits_0^{1 - x}\!\d\,y\,\Bigl[\frac{x}{{\cal{\tM}}^2} \Bigl\{\frac{y^2 p_m p_n}{8{\cal{\tM}}^2} - \eta_{mn}\Bigl\}\Bigl]
\labl{B24}
\end{eqnarray}
\setcounter{equation}{0}

\bibliographystyle{paper}
{\small
\bibliography{paper}
}

\end{document}

%% file: texdefs.tex
\renewcommand{\d}{\mathrm{d}}

\newcommand{\captn}[1]{\vspace{-3ex}\caption{\small #1}}

\newcommand{\dga}{{\dot{\alpha}}}
\newcommand{\dgb}{{\dot{\beta}}}
\newcommand{\dgg}{{\dot{\gamma}}}

\newcommand{\ga}{\alpha}
\newcommand{\gb}{\beta}
\renewcommand{\gg}{\gamma}
\newcommand{\gd}{\delta}
\renewcommand{\ge}{\epsilon}

\newcommand{\gm}{\mu}

\newcommand{\gl}{\lambda}

\newcommand{\gth}{\theta}

\newcommand{\gs}{\sigma}

\newcommand{\gp}{\pi}
\newcommand{\gps}{\psi}
\newcommand{\get}{\eta}

\newcommand{\gG}{\Gamma}
\newcommand{\gD}{\Delta}
\newcommand{\gF}{\Phi}

\newcommand{\gS}{\Sigma}
\newcommand{\gTh}{\Theta}

\newcommand{\gPs}{\Psi}

\newcommand{\cD}{{\cal D}}

\newcommand{\tm}{{\widetilde m}}

\newcommand{\tq}{{\widetilde q}}

\newcommand{\tM}{{\widetilde M}}

\newcommand{\Id}{\text{\small 1}\hspace{-3.5pt}\text{1}}
\newcommand{\slashed}{\hspace{-1.1ex}/}

\newcommand{\der}{\partial}

\newcommand{\inv}{^{-1}}
\newcommand{\dsp}{\displaystyle}

\newcommand{\labl}[1]{\label{#1}}
\newcommand{\Kh}{K\"{a}hler}
\newcommand{\beq}{\begin{equation}}
\newcommand{\eeq}{\end{equation}}
\newcommand{\barr}{\begin{array}}
\newcommand{\earr}{\end{array}}
\newcommand{\equ}[1]{\begin{gather} #1 \end{gather}}
\newcommand{\equa}[1]{\begin{align} #1 \end{align}}

\newcommand{\tabu}[2]{\begin{tabular}{#1} #2 \end{tabular}}
\newcommand{\arry}[2]{\begin{array}{#1} #2 \end{array}}

\newcommand{\pmtrx}[1]{\begin{pmatrix} #1 \end{pmatrix}}

\newcommand{\non}{\nonumber}
\newcounter{oldcounter}

\newcommand{\bC}{{\overline C}}
\newcommand{\bD}{{\overline D}}

\newcommand{\bF}{{\overline F}}

\newcommand{\bJ}{{\overline J}}

\newcommand{\bW}{{\overline W}}

\newcommand{\bge}{{\bar\epsilon}}

\newcommand{\bgm}{{\bar\mu}}

\newcommand{\bgl}{{\bar\lambda}}

\newcommand{\bgth}{{\bar\theta}}
\newcommand{\bgs}{{\bar\sigma}}

\newcommand{\bgps}{{\bar\psi}}
\newcommand{\bget}{{\bar\eta}}

\newcommand{\bgF}{{\overline\Phi}}

\newcommand{\bgTh}{{\overline\Theta}}

\newcommand{\bgPs}{{\overline\Psi}}

\newcommand{\tgg}{{\widetilde \gamma}}

\newcommand{\ba}[2]{\[\begin{array}{#2}\label{#1}}
\newcommand{\ea}{\end{array}\]}
\newcommand{\be}{\begin{equation}}
\newcommand{\ee}{\end{equation}}
\newcommand{\bea}{\begin{eqnarray}}
\newcommand{\eea}{\end{eqnarray}}

\newcommand{\U}[1]{\mathrm{U(#1)}}

\newcommand{\MSbar}{$\overline{\mbox{MS}}$}

%% file: propbgFgF.pstex_t
\begin{picture}(0,0)%
\includegraphics{propbgFgF.pstex}%
\end{picture}%
\setlength{\unitlength}{4144sp}%
\begingroup\makeatletter\ifx\SetFigFont\undefined%
\gdef\SetFigFont#1#2#3#4#5{%
  \reset@font\fontsize{#1}{#2pt}%
  \fontfamily{#3}\fontseries{#4}\fontshape{#5}%
  \selectfont}%
\fi\endgroup%
\begin{picture}(1844,314)(2004,-3818)
\end{picture}

%% file: propgFgF.pstex_t
\begin{picture}(0,0)%
\includegraphics{propgFgF.pstex}%
\end{picture}%
\setlength{\unitlength}{4144sp}%
\begingroup\makeatletter\ifx\SetFigFont\undefined%
\gdef\SetFigFont#1#2#3#4#5{%
  \reset@font\fontsize{#1}{#2pt}%
  \fontfamily{#3}\fontseries{#4}\fontshape{#5}%
  \selectfont}%
\fi\endgroup%
\begin{picture}(1844,314)(2004,-3818)
\end{picture}

%% file: propVV.pstex_t
\begin{picture}(0,0)%
\includegraphics{propVV.pstex}%
\end{picture}%
\setlength{\unitlength}{4144sp}%
\begingroup\makeatletter\ifx\SetFigFont\undefined%
\gdef\SetFigFont#1#2#3#4#5{%
  \reset@font\fontsize{#1}{#2pt}%
  \fontfamily{#3}\fontseries{#4}\fontshape{#5}%
  \selectfont}%
\fi\endgroup%
\begin{picture}(1844,318)(1959,-3685)
\end{picture}

%% file: chiral_self.pstex_t
\begin{picture}(0,0)%
\includegraphics{chiral_self.pstex}%
\end{picture}%
\setlength{\unitlength}{4144sp}%
\begingroup\makeatletter\ifx\SetFigFont\undefined%
\gdef\SetFigFont#1#2#3#4#5{%
  \reset@font\fontsize{#1}{#2pt}%
  \fontfamily{#3}\fontseries{#4}\fontshape{#5}%
  \selectfont}%
\fi\endgroup%
\begin{picture}(3624,727)(2014,-4916)
\end{picture}%

%% file: chiral_tad.pstex_t
\begin{picture}(0,0)%
\includegraphics{chiral_tad.pstex}%
\end{picture}%
\setlength{\unitlength}{4144sp}%
\begingroup\makeatletter\ifx\SetFigFont\undefined%
\gdef\SetFigFont#1#2#3#4#5{%
  \reset@font\fontsize{#1}{#2pt}%
  \fontfamily{#3}\fontseries{#4}\fontshape{#5}%
  \selectfont}%
\fi\endgroup%
\begin{picture}(3624,710)(2014,-3943)
\end{picture}%